\documentclass[10pt,conference,onecolumn]{IEEEtran}

\usepackage{graphicx}
\usepackage{amssymb,amsmath}
\usepackage{verbatim}
\usepackage{subfigure}

\newtheorem{lem}{Lemma}

\newtheorem{thm}{Theorem}
\newtheorem{defn}{Definition}
\newtheorem{remark}{Remark}
\newtheorem{algorithm}{Algorithm}

\begin{document}

\title{Optimal Routing for the Gaussian Multiple-Relay Channel with Decode-and-Forward}

\author{\authorblockN{Lawrence Ong and Mehul Motani}
\authorblockA{Department of Electrical and Computer Engineering\\
National University of Singapore\\
Email: \{lawrence.ong,motani\}@nus.edu.sg\\}
}

\maketitle

\begin{abstract}
In this paper, we study a routing problem on the Gaussian multiple relay channel, in which nodes employ a decode-and-forward coding strategy.  We are interested in routes for the information flow through the relays that achieve the highest DF rate. We first construct an algorithm that provably finds optimal DF routes. As the algorithm runs in factorial time in the worst case, we propose a polynomial time heuristic algorithm that finds an optimal route with high probability.  We demonstrate that that the optimal (and near optimal) DF routes are good in practice by simulating  a distributed DF coding scheme using low density parity check codes with puncturing and incremental redundancy.

\end{abstract}


%

\section{Introduction}


The wireless channel allows intermediate nodes in a network to overhear the transmissions of a source-destination pair and act as relays.  We take 
an information theoretic approach to understand communication in this network scenario.
The single-relay channel (SRC) was introduced by van der Meulen~\cite{meulen71}. To date, the largest achievable region for the general SRC is due to Cover and El Gamal~\cite{covergamal79}, who constructed two coding strategies, commonly referred to as \emph{decode-and-forward} (DF) and \emph{compress-and-forward} (CF).  Chong, Motani, \& Garg ~\cite{chongmotani07} recently introduced a different decoding technique to give a potentially larger achievable region for the general SRC. The SRC was extended to the multiple-relay channel (MRC) by Gupta and Kumar~\cite{guptakumar03}, and Xie and Kumar~\cite{xiekumar03}, who presented an achievable rate region based on DF. The capacity of the MRC is not known except for special cases, including the degraded MRC~\cite{xiekumar03} (achievable by DF), the phase fading MRC where the relays are within a certain distance from the source~\cite{kramergastpar04} (achievable by DF), and mesh networks~\cite{ongmotani06a}\cite{ongmotani07} (achievable by CF).

In DF, ``\emph{we can imagine that there is an information flow from the source node... to the destination node...}'' \cite{xiekumar03}. We call this flow a \emph{route}. In this paper, we construct an algorithm to find the optimal (i.e., rate maximizing) route for DF on the Gaussian MRC. However, we note that the algorithm runs in factorial time in the worst case. We exploit properties of the algorithm to design a heuristic algorithm that runs in polynomial time and finds optimal routes with high probability.

Recently,  low-density parity-check (LDPC) codes~\cite{razaghiyu06}\cite{ezrigastpar06}\cite{khojastepouraazhang04}\cite{chakrabarti06} and Turbo codes~\cite{zhaovalenti03}\cite{zhangbahceci04} for the SRC have been constructed based on DF. In this paper, we simulate the DF-based coding strategy using LDPC codes \cite{gallager62}\cite{mackay99} with incremental redundancy on the MRC and compare the performance of different routes.

Our contributions in this paper are as follows.
\begin{enumerate}
\item We construct an algorithm that finds optimal (rate maximizing) routes for DF on the Gaussian MRC.
\item We construct a heuristic algorithm that runs in polynomial time and finds optimal routes with high probability. 
\item We implement DF on the MRC using LDPC codes \cite{gallager62}\cite{mackay99} and demonstrate that the DF optimal routes are good in practice.
\end{enumerate}

\section{Network Model}\label{sec:routing_channel_model}
We consider a $D$-node Gaussian MRC: $\mathcal{S} =\{1, 2, 3 \dotsc,$ $D-1, D\}$ with one source (node 1) and one destination (node $D$). 
We use the standard path loss model for signal propagation. The received signal at node $t$, $t=2, 3, \dotsc, D$, can be written as
$Y_t = \sum_{\substack{i=1,i\neq t}}^{D-1} \sqrt{\kappa d_{it}^{-\eta}} X_i + Z_t$,
where the transmit signal of node $i$, $X_i$, is a random variable with average power constraint $E[X_i^2] \leq P_i$. $d_{it}$ is the distance between nodes $i$ and $t$, $\eta$ is the path loss exponent
($\eta \geq 2$ with equality for free space transmission), and $\kappa$ is a positive constant.
$Z_t$ is the receiver noise at node $t$, which is an independent, zero mean Gaussian random variable with variance $N_t$.
We assume duplex nodes and that all nodes have the same noise variance.

\subsection{Route}
Now, we define what we mean by a route in a network. 
\begin{defn}
The route taken by a packet from the source to the destination is an ordered set of nodes involved in encoding/transmitting of the packet. The sequence of the nodes in the route is determined by the order in which nodes' transmit signals first depend on the packet.  Note that the destination node is the last node in the route and does not transmit the data.
\end{defn}

\begin{remark}
For DF, a route is also the order in which a message is fully decoded at the nodes.
\end{remark}

In the rest of this paper, we denote a route by $\mathcal{M} = \{ m_1, m_2, \dotsc, m_{|\mathcal{M}|} \}$.
We define the set of all possible routes from the source (node 1) to the destination (node $D$) by $\Pi (\mathcal{S}) = \Big\{ \{ m_1, m_2, \dotsc, m_{|\mathcal{M}|}\}: m_2,\dotsc,m_{|\mathcal{M}|-1}$ are all possible selections and permutations of the relays (including the empty set), $m_1=1, m_{|\mathcal{M}|}=D \Big\}$.

It has been noted~\cite{xiekumar03}\cite{kramergastpar04} that the DF rate depends on the route selected (known as node order/permutation in the papers). Besides the \emph{degraded} case, finding the optimal route is not easy in general. We refer to the strategy of testing all possible routes as \emph{brute force}.

\subsection{Achievable Rates using DF}
Using DF~\cite{xiekumar03}\cite{kramergastpar04}\cite{xiekumar04} with Gaussian inputs and route $\mathcal{M}$, node $m_i$ transmits $X_{m_i} = \sum_{j=i+1}^{|\mathcal{M}|} \sqrt{\alpha_{m_im_j}P_{m_i}}U_{m_j}$,
for $0 \leq \sum_{j=i+1}^{|\mathcal{M}|} \alpha_{m_i m_j} \leq 1$, $\forall i=1, \dotsc, |\mathcal{M}|-1$. $U_{m_j}$ are independent Gaussian random variables with unit variance. $\{\alpha_{m_im_j}|j=i+1, \dotsc, |\mathcal{M}|\}$ are the power splits of node $m_i$, allocating portions of its transmit power to transmit independent \emph{sub-codewords} $U_{m_j}$.

On route $\mathcal{M}$, DF can achieve rates up to
\begin{equation}\label{eq:df-rate}
R_{\text{DF}}(\mathcal{M}) = \max_{\alpha_{ij}} \min_{ m_t \in \mathcal{M} \setminus \{ m_1 \}} R_{m_t}(\mathcal{M}),
\end{equation}
where $R_{m_t}(\mathcal{M})$ is the \emph{reception rate} of the node $m_t$ in route $\mathcal{M}$, given by
\begin{equation}
R_{m_t}(\mathcal{M}) = L \left( N_{m_t}^{-1}\sum_{j=2}^t \left( \sum_{i=1}^{j-1}\sqrt{ \alpha_{m_im_j}P_{m_im_t}} \right)^2 \right),
\end{equation}
where $P_{m_im_t} = \kappa d_{m_im_t}^{-\eta}P_{m_i}$ and $L(x) = \frac{1}{2}\log(1 + x)$.
We call the rate in \eqref{eq:df-rate} the DF rate \emph{supported} by route $\mathcal{M}$. The maximum DF rate  is
\begin{equation} \label{eq:max-df-rate}
R_{\text{DF}}^{\text{max}} = \max_{\mathcal{M} \in \Pi(\mathcal{S})} R_{\text{DF}}(\mathcal{M}).
\end{equation}

\subsection{The Optimal Routing Problem}
We define the optimal route set for DF as
\begin{equation}
\mathcal{Q}_{\text{DF}} \triangleq \{\mathcal{M}\in\Pi(\mathcal{S}) | R_{\text{DF}}(\mathcal{M})=R_{\text{DF}}^{\text{max}}\}\nonumber,
\end{equation}
We define the optimal route set because the rate maximizing route may not be unique. Then the optimal DF routing problem for a network $\mathcal{S}$ is to find $\mathcal{M}_{\text{DF}}^{\text{opt}} \in \mathcal{Q}_{\text{DF}}$.

\section{The Nearest Neighbor Algorithm}\label{sec:routing_nna}
In this section, we present an algorithm to find an optimal DF route.  First, we define the \emph{nearest neighbor} with respect to a route.
\begin{defn}
Node $i \notin \mathcal{M}$ is a nearest neighbor with respect to the route $\mathcal{M}$ iff
\begin{equation}\label{eq:nearest_neighbor_condition}
P_{mi} \geq P_{mj}, \quad \forall m \in \mathcal{M}, \forall j \in \mathcal{S} \setminus (\mathcal{M} \cup \{ i \}).
\end{equation}
\end{defn}
Now, we describe the nearest neighbor algorithm (NNA).
\begin{algorithm}[NNA]
\mbox{\,}
\begin{enumerate}
\item First, start with the source node, $\mathcal{M} = \{ m_1\}$.
\item \label{item:pick_neighbor} 
If there exists a unique nearest neighbor $i^*$ with respect to the current route $\mathcal{M}$, we append $i^*$ to the current route: 
$\mathcal{M} \leftarrow \mathcal{M} \cup \{ i^* \}$. Else, the NNA terminates prematurely. Since $\mathcal{M}$ is an ordered set, the notation $\mathcal{A} \cup \mathcal{B}$ means appending ordered set $\mathcal{B}$ to the end of ordered set $\mathcal{A}$.
\item Step~\ref{item:pick_neighbor} is repeated until the destination node, node $D$, is added into $\mathcal{M}$.
\end{enumerate}
\end{algorithm}

The algorithm is said to terminate normally if node $D$ is added to the route. Otherwise, the algorithm is said to terminate prematurely. If the NNA terminates normally, we have the following theorem.

\begin{thm} \label{thm:NNA}
Consider a Gaussian MRC. If the NNA terminates normally, then the NNA route is optimal for DF.
\end{thm}

To prove Theorem~\ref{thm:NNA}, we need the following lemmas.

\begin{lem}\label{lem:nearest_neighbor}
When we add the unique nearest neighbor, node $a^*$, to route $\mathcal{M}$, the rate supported by the new route $\mathcal{M}_1 = \mathcal{M} \cup \{m_{|\mathcal{M}|+1}=a^*\}$ is greater than or equal to the rate supported by the route formed by adding any other node to $\mathcal{M}$, $\mathcal{M}_2 = \mathcal{M} \cup \{m_{|\mathcal{M}|+1}=b\}$.
Mathematically,
\begin{equation}
R_{\text{DF}}(\mathcal{M} \cup \{a^*\}) \geq R_{\text{DF}}(\mathcal{M} \cup \{b\}), \forall b \in \mathcal{S} \setminus ( \mathcal{M} \cup \{a^*\} ).
\end{equation}
\end{lem}

\begin{proof}[Proof for Lemma~\ref{lem:nearest_neighbor}] Considering $\mathcal{M}_2$, the reception rate at node $m_{|\mathcal{M}|+1}=b$ is
\begin{equation}
R_b(\mathcal{M}_2) = L \left( N_b^{-1} \sum_{j=2}^{|\mathcal{M}|+1} \left( \sum_{i=1}^{j-1} \sqrt{\alpha_{m_i m_j}^{(2)}P_{m_ib}} \right)^2 \right).
\end{equation}
Here, we choose the set of $\{\alpha_{ij}^{(2)}\}$ that maximizes $\min_{2 \leq i \leq |\mathcal{M}|+1} R_{m_i}(\mathcal{M}_2)$. Using the same set of $\{\alpha_{ij}^{(2)}\}$, the reception rate of the node $m_{|\mathcal{M}|+1}=a^*$ in route $\mathcal{M}_1$ is
\begin{equation}
R_{a^*}(\mathcal{M}_1) = L \left( N_{a^*}^{-1}\sum_{j=2}^{|\mathcal{M}|+1} \left( \sum_{i=1}^{j-1} \sqrt{\alpha_{m_i m_j}^{(2)}P_{m_ia^*}} \right)^2 \right).
\end{equation}
Note that this set of $\{\alpha_{ij}^{(2)}\}$ does not necessarily maximizes $\min_{2 \leq i \leq |\mathcal{M}|+1} R_{m_i}(\mathcal{M}_1)$. Clearly, if $P_{m a^*} \geq P_{m b}, \quad \forall m \in \mathcal{M}$ with at least one inequality and $N_{a^*} = N_b$, we get $R_{a^*}(\mathcal{M}_1) \geq R_b(\mathcal{M}_2)$.

Using $\{\alpha_{ij}^{(2)}\}$, the reception rates of the first $|\mathcal{M}|$ nodes remain the same for both $\mathcal{M}_1$ and $\mathcal{M}_2$. Using the power split optimized for $\mathcal{M}_1$, we might be able to increase the minimum reception rate in  $\mathcal{M}_1$.  Hence $R_{\text{DF}}(\mathcal{M}_1) \geq R_{\text{DF}} (\mathcal{M}_2)$.
\end{proof}

Next, we show that choosing the nearest neighbor will not harm the rate supported by the route even when more nodes are added.

\begin{lem}\label{lem:nearest_neighbor_in_middle}
Let $\mathcal{M} = \{ a_1^*, a_2^*, \dotsc, a_{|\mathcal{M}|}^*\}$
be a route formed by adding the nearest neighbor (assuming it exists) one by one starting from the source. Now, arbitrarily add $K$ nodes to $\mathcal{M}$. The first node $b_1$ is not a nearest neighbor and the rest may or may not be nearest neighbors. In other words,
$
\mathcal{M}_1 = \{ a_1^*, a_2^*, \dotsc, a_{|\mathcal{M}|}^*, b_1, b_2, \dotsc, b_K \},
$
where $b_1$ is not a nearest neighbor to $\mathcal{M}$. We can always choose the nearest neighbor $a_{{|\mathcal{M}|}+1}^*$ (assuming it exists) at the $(|\mathcal{M}|+1)$-th position, i.e.,
\begin{equation} \label{eq:lemma-2-eq}  \scriptstyle
\mathcal{M}_2 = 
\begin{cases}
\{ a_1^*, \dotsc, a_{|\mathcal{M}|}^*, a_{{|\mathcal{M}|}+1}^*, b_1, \dotsc, b_{K-1} \},\\
\quad\quad \text{if } a_{{|\mathcal{M}|}+1}^* \notin \{b_1, \dotsc, b_{K-1} \}, \\
\{ a_1^*, \dotsc, a_{|\mathcal{M}|}^*, a_{{|\mathcal{M}|}+1}^*, b_1, \dotsc, b_{k-1}, b_{k+1}, \dotsc, b_K \},\\
\quad\quad \text{if $a_{{|\mathcal{M}|}+1}^* = b_k$, for some $b_k \in \{b_1, \dotsc, b_{K-1} \}$}
\end{cases}
\end{equation}
and show that
$
R_{\text{DF}}(\mathcal{M}_2) \geq R_{\text{DF}}(\mathcal{M}_1).
$
\end{lem}

\begin{proof}[Proof for Lemma~\ref{lem:nearest_neighbor_in_middle}] First, we study the case when $a_{|\mathcal{M}|+1}^* \notin \{b_1, \dotsc, b_{K-1} \}$. Let the optimum set of power splits for $\mathcal{M}_1$ be $\{\alpha_{ij}^{(1)}\}$.
Using this set $\{\alpha_{ij}^{(1)}\}$ for the nodes in $\mathcal{M}_2$ and some $\{\alpha_{a_{|\mathcal{M}|+1}^*j}'\}$ for node $a_{|\mathcal{M}|+1}^*$,
\begin{subequations}
\begin{align}
R_{a_{|\mathcal{M}|+1}^*}(\mathcal{M}_2) & \geq R_{b_1}(\mathcal{M}_1) \label{eq:explain3a}\\
R_{a_i^*}(\mathcal{M}_2) & =  R_{a_i^*}(\mathcal{M}_1), \quad \forall i = 2, 3, \dotsc , |\mathcal{M}| \label{eq:explain3} \\
R_{b_i}(\mathcal{M}_2) & \geq R_{b_i}(\mathcal{M}_1), \quad \forall i = 1, 2, \dotsc , K-1. \label{eq:explain4}
\end{align}
\end{subequations}
Equation~\eqref{eq:explain3a} is due to Lemma~\ref{lem:nearest_neighbor}.
Equation~\eqref{eq:explain3} is because nodes transmitting to $a_i^*$ in both routes uses the same power split. Equation~\eqref{eq:explain4} is because all the nodes except $a_{|\mathcal{M}|+1}^*$ uses the same power split for both routes and with an additional node $a_{|\mathcal{M}|+1}^*$ transmitting to nodes $\{b_1, \dotsc, b_{K-1} \}$ in $\mathcal{M}_2$, there is a possible increase in the reception rate.

Now, we study the case when $a_{|\mathcal{M}|+1}^* = b_k$ for some $b_k \in \{b_1, \dotsc, b_{K-1} \}$. Now, using $\{\alpha_{ij}^{(1)}\}$ for all the nodes in $\mathcal{M}_2$,
\begin{subequations}
\begin{align}
R_{a_i^*}(\mathcal{M}_2) & =  R_{a_i^*}(\mathcal{M}_1), \quad \forall i = 2, 3, \dotsc , |\mathcal{M}| \label{eq:explain5} \\
R_{a_{|\mathcal{M}|+1}^*}(\mathcal{M}_2) & \geq R_{b_1}(\mathcal{M}_1), \label{eq:explain6} \\
R_{b_i}(\mathcal{M}_2) & = R_{b_i}(\mathcal{M}_1), \quad \forall i = 1, \dotsc , k-1, k+1, \dotsc, K. \label{eq:explain7}
\end{align}
\end{subequations}
Equations~\eqref{eq:explain5} and \eqref{eq:explain7} are because nodes transmit using the same power split  $\{\alpha_{ij}^{(1)}\}$ in both routes. Equation~\eqref{eq:explain6} follows from Lemma~\ref{lem:nearest_neighbor}.

Now, let $\{\alpha_{ij}^{(2)}\}$ be a set of power splits optimized for $\mathcal{M}_2$.
\begin{subequations}
\begin{align}
& R_{\text{DF}}(\mathcal{M}_2)\\
& = \min_{k \in \mathcal{M}_2 \setminus \{ a^*_1\}}  R_k(\mathcal{M}_2), \quad \text{using $\{\alpha_{ij}^{(2)}\}$} \\
& \geq \min_{k \in \mathcal{M}_2 \setminus \{ a^*_1\}} R_k(\mathcal{M}_2), \quad \text{using $\{\alpha_{ij}^{(1)}\}$, $\{\alpha_{a_{|\mathcal{M}|+1}^*j}'\}$} \\
& \geq \min_{k \in \mathcal{M}_1 \setminus \{ a^*_1\}} R_k(\mathcal{M}_1), \quad \text{using $\{\alpha_{ij}^{(1)}\}$} \\
& = R_{\text{DF}}(\mathcal{M}_1).
\end{align}
\end{subequations}
This proves Lemma~\ref{lem:nearest_neighbor_in_middle}.
\end{proof}

\begin{lem}\label{lem::same_length_nna_best}
The DF rate supported by a route that contains all nearest neighbors is always higher or equal to that by any route of the same length, with non-nearest neighbor(s).
\end{lem}

\begin{proof}[Proof for Lemma~\ref{lem::same_length_nna_best}]
Lemma~\ref{lem::same_length_nna_best} can be proven by applying Lemma~\ref{lem:nearest_neighbor_in_middle} recursively until the entire route is replaced by nearest neighbor nodes. 
\end{proof}

Now we consider routes from the source to the destination, with possibly different lengths.

\begin{proof}[Proof for Theorem~\ref{thm:NNA}]
Consider a route $\mathcal{M}_1$ from the source to the destination, which contains one or more non-nearest-neighbor(s).
If NNA terminates normally and outputs $\mathcal{M}_2$, we want to show that $R_{\text{DF}}(\mathcal{M}_2) \geq R_{\text{DF}}(\mathcal{M}_1)$.
We note that ${|{M}_1|}$ does not necessarily equal ${|{M}_2|}$. 

Let $\mathcal{M}_1 = \{ m_1^*=1, m_2, \dotsc, m_{|{M}_1|}=D \}$ and
$\mathcal{M}_2 = \{ m_1^*=1, m_2^*, \dotsc, m_{|{M}_2|}^*=D \}$. We use asterisks to mark nearest neighbor nodes.

First of all, we consider the case ${|{M}_1|}={|{M}_2|}$. The results follows immediately from Lemma~\ref{lem::same_length_nna_best}. Second, we consider ${|{M}_1|} > {|{M}_2|}$. We consider first ${|{M}_2|}$ nodes in $\mathcal{M}_1$, i.e., 
$\mathcal{M}_1' = \{ m_1^*, m_2, \dotsc, m_{|{M}_2|}\}$.
Then,
\begin{equation}
R_{\text{DF}}(\mathcal{M}_2) \geq R_{\text{DF}}(\mathcal{M}_1') \geq R_{\text{DF}}(\mathcal{M}_1).
\end{equation}
The first inequality is obtained by applying Lemma~\ref{lem::same_length_nna_best}, as $|\mathcal{M}_2| = |\mathcal{M}_1'|$. The second inequality can be argued as follows. The first $|\mathcal{M}_2|$ nodes in both routes $\mathcal{M}_1'$ and $\mathcal{M}_1$ are identical. Hence the reception rates are the same using the same power split. However, there are additional nodes in $\mathcal{M}_1$ whose reception rate might be lower than $R_{\text{DF}}(\mathcal{M}_1')$. Hence, $R_{\text{DF}}(\mathcal{M}_1) \leq R_{\text{DF}}(\mathcal{M}_1')$.

Lastly, consider ${|{M}_2|} > {|{M}_1|}$. We replace the transmitting nodes in $\mathcal{M}_1$ with nearest neighbors and obtain
\begin{equation}
\mathcal{M}_3 = \{ m_1^*, m_2^*, \dotsc, m_{|{M}_1|-1}^*, m_{|{M}_1|}=D \}.
\end{equation}
Note that $m_{|{M}_1|}$ is not be the nearest neighbor. Clearly, using Lemma~\ref{lem:nearest_neighbor_in_middle},
$R_{\text{DF}}(\mathcal{M}_3) \geq R_{\text{DF}}(\mathcal{M}_1)$.

Now, using the set of power splits $\{\alpha_{ij}^{(3)}\}$ optimized for $\mathcal{M}_3$, we have
\begin{subequations}
\begin{align}
& R_{\text{DF}}(\mathcal{M}_2)\\
& = \min_{k \in \mathcal{M}_2 \setminus \{ a^*_1\}} R_{k}(\mathcal{M}_2), \quad \text{using $\{\alpha_{ij}^{(2)}\}$} \\
& \geq \min_{k \in \mathcal{M}_2 \setminus \{ a^*_1\}} R_{k}(\mathcal{M}_2), \quad \text{using $\{\alpha_{ij}^{(3)}\}$  and some $\{\alpha_{ij}\}$ for} \nonumber \\
& \quad \text{nodes $m_{|\mathcal{M}_1|}^*, \dotsc, m_{|\mathcal{M}_2|-1}^*$} \\
& \geq \min_{k \in \mathcal{M}_3 \setminus \{ a^*_1\}} R_{k}(\mathcal{M}_3), \quad \text{using $\{\alpha_{ij}^{(3)}\}$} \label{lemma4-1} \\
& = R_{\text{DF}}(\mathcal{M}_3) \geq R_{\text{DF}}(\mathcal{M}_1).
\end{align}
\end{subequations}

The inequality in \eqref{lemma4-1} is because using the NNA, $\{ m^*_{|\mathcal{M}_1|}, \dotsc, m^*_{|\mathcal{M}_2|-1} \}$ are added to $\{m_1^*, \dotsc, m^*_{|\mathcal{M}_1|-1}\}$ before $D$. A necessary condition for this is
\begin{equation}
P_{m n} \geq P_{m D},\forall m \in  \{m_1^*, \dotsc, m^*_{|\mathcal{M}_1|-1}\}, \forall n \in \{ m^*_{|\mathcal{M}_1|}, \dotsc, m^*_{|\mathcal{M}_2|-1} \},
\end{equation}
Hence,
\begin{equation}
R_n(\mathcal{M}_2) \geq R_D(\mathcal{M}_3), \forall n \in \{ m^*_{|\mathcal{M}_1|}, \dotsc, m^*_{|\mathcal{M}_2|-1} \}.
\end{equation}
With additional nodes transmitting to $D$ in $\mathcal{M}_2$,
$R_D(\mathcal{M}_2) > R_D(\mathcal{M}_3)$.
Hence, we have Theorem~\ref{thm:NNA}.
\end{proof}

\begin{remark}
We note the NNA terminates normally if and only if a unique nearest neighbor exists at each step. In the next section, we extend the NNA to an algorithm which terminates normally given any network topology.
\end{remark}

\section{The Nearest Neighbor Set Algorithm}\label{sec:routing_nnsa}
In this section, we modify the NNA so that it terminates normally in any Gaussian MRC. We term this algorithm the nearest neighbor set algorithm (NNSA). First, we define the \emph{nearest neighbor set}.
\begin{defn}
The nearest neighbor set $\mathcal{N} = \{ n_1, n_2, \dotsc, n_{|{N}|} \}$ with respect to route $\mathcal{M} = \{ m_1, m_2,$ $\dotsc, m_{|\mathcal{M}|} \}$ is defined as the smallest set $\mathcal{N}$ where each $n \in \mathcal{N} \subseteq \mathcal{S} \setminus \mathcal{M}$ satisfies the following condition.
\begin{equation}
P_{mn} \geq P_{ma}, \quad \forall m \in \mathcal{M}, \forall a \in \mathcal{S} \setminus( \mathcal{M} \cup \mathcal{N}),
\end{equation}
with at least one strict inequality for every pair of $(n,a) \in \{ (n,a) \vert n \in \mathcal{N}, a \in \mathcal{S} \setminus (\mathcal{M} \cup \mathcal{N}) \}$.
\end{defn}

Now we describe the NNSA.
\begin{algorithm}[NNSA]
\mbox{\,}
\begin{enumerate}
\item Starting with the source node, we have $\mathcal{M} = \{ 1 \}$.
\item \label{item:NNSA_add_nodes}
Find the nearest neighbor set $\mathcal{N}$. 
The original route $\mathcal{M}$ branches out to $|\mathcal{N}|$ new routes as follows:
\begin{equation}
\mathcal{M}_i \leftarrow \mathcal{M} \cup \{ n_i \}, \quad i=1, \dotsc, |\mathcal{N}|. \label{eq:NNSA}
\end{equation}
\item For each new route in \eqref{eq:NNSA}, step~\ref{item:NNSA_add_nodes} is repeated until the destination is added to all routes.
\end{enumerate}
\end{algorithm}

When the algorithm terminates, we end up with many routes from the source to the destination. We term these routes \emph{NNSA candidates}. We calculate the supported rate of each candidate and choose the one which gives the highest supported rate. The following theorem says that the NNSA candidate that gives the highest supported rate is an optimal route for DF.

\begin{thm} \label{thm:NNSA}
Consider a Gaussian MRC. The NNSA candidate routes that give the highest supported rate are optimal for DF.
\end{thm}

\begin{proof}[Sketch of proof for Theorem~\ref{thm:NNSA}] Using the technique used in the proof of Theorem~\ref{thm:NNA}, we can show that adding a node that does not belong to the nearest neighbor set can only be suboptimal. We can always replace that node with one from the nearest neighbor set and obtain an equal or higher rate. In other words, consider a non-NNSA candidate
$\mathcal{M}_1 = \{ m_1^*=1, m_2, \dotsc, m_{|{M}_1|}=D \}$, where one or more nodes in $\{m_2, \dotsc, m_{|{M}_1|}\}$ are not from the nearest neighbor set, and an NNSA candidate $\mathcal{M}_2 = \{ m_1^*=1, m_2^*, \dotsc, m_{|{M}_2|}^*=D\}$,
where all nodes in $\mathcal{M}_2$ are added according to the NNSA. We can show that
$R_{\text{DF}}(\mathcal{M}_2) \geq R_{\text{DF}}(\mathcal{M}_1)$.

The NNSA finds all possible routes for which every node is added from the nearest neighbor set. Hence one or more of the NNSA candidate must achieve the highest DF rate.
This gives us Theorem~\ref{thm:NNSA}.
\end{proof}

\begin{remark}
We can show that a shortest optimal route, $\mathcal{M}_{\text{DF}}^{\text{SOR}} \in \mathcal{Q}_{\text{DF}}$, s.t. $\lvert \mathcal{M}_{\text{DF}}^{\text{SOR}} \rvert \leq |\mathcal{M}|, \forall \mathcal{M} \in \mathcal{Q}_{\text{DF}}$, is contained in one of the NNSA candidates that supports $R_{\text{DF}}^{\text{max}}$.
\end{remark}

\begin{remark} \label{rem:ic}
We note that the NNSA is also optimum in the phase fading Gaussian MRC, where all node transmit independent codewords, i.e., $\alpha_{ij}=1, \forall i \in \mathcal{M}\setminus\{ D \}, j=i+1$ and $\alpha_{ij}=0, \forall j \neq i+1$.
\end{remark}


\section{Complexity of the NNSA} \label{sec:routing_simulation}
With the NNSA, we can now search for the optimal route in the NNSA candidate set, as compared to searching in $\Pi(\mathcal{S})$ using brute force. The number of candidates determines the number of routes whose rate we need to calculate to find optimal routes. We note that the size of the NNSA candidate set might still, in the worst case, equal $|\Pi(\mathcal{S})|$. Using brute force, the number of permutations we need to check is
\begin{equation}
|\Pi(\mathcal{S})| = 1 + \binom{D-2}{1} + \binom{D-2}{2} + \dotsm \binom{D-2}{D-2}
 = O((D-1)!),
\end{equation}
where $D$ is the total number of nodes in the network and $\binom{n}{k} = \frac{n \times (n-1) \times \dotsm \times 1}{(n-k) \times (n-k-1) \times \dotsm \times 1}$.

We ran the NNSA on 10000 randomly generated networks with a varying number of nodes uniformly distributed in a 1m$\times$1m square area. The source, relays, and the destination were randomly assigned. On average, half of the NNSA candidate set sizes were less than 0.715\% of $|\Pi(\mathcal{S})|$ for the 8-node channel and less than 0.253\% of $|\Pi(\mathcal{S})|$ for the 11-node channel.

We note that the average size of the NNSA candidate set does grow factorially with the number of nodes. However this does increase the range of finite size networks for which we can find optimal routes.
Furthermore, the NNSA provides insights for designing heuristic algorithms to find good routes for DF-based codes. In the next section, we propose a heuristic algorithm which outputs routes in polynomial time.

\section{A Heuristic Algorithm} \label{sec:NWCA}
In the NNSA, the optimal route is constructed by adding the ``next hop'' node one by one to the \emph{partial} route. The node to be added is from the nearest neighbor set. If the nearest neighbor set contains more than one node, the current route branches to more than one routes, leading to a possibly large NNSA candidate set size. 

To avoid this, we consider a heuristic approach that starts from the source node and repeatedly adds only one ``good'' candidate from the nearest neighbor set until the destination is reached. For the choice of the next hop node, we consider the node which receives the largest sum of received power from all the nodes in the existing partial route. We call this the maximum sum-of-received-power algorithm (MSPA).  By choosing only one node to be added to the partial route, we prevent the algorithm from branching out to multiple routes. This heuristic approach yields only one route, regardless of the network size.  We now explicitly describe the MSPA. 

\begin{algorithm}[MSPA]
\mbox{\,}
\begin{enumerate}
\item First, start with the source node, $\mathcal{M} = \{ m_1\}$.
\item \label{item:mspa1} For every node $t \in \mathcal{S} \setminus \mathcal{M}$, find the sum of received power from all nodes in $\mathcal{M}$ to $t$, $\sum_{i \in \mathcal{M}} P_{it}$.
\item \label{item:mspa2} Let $a^*$ be {\em any} node with the highest sum of received power, i.e., $\sum_{i \in \mathcal{M}} P_{ia^*} \geq \sum_{j \in \mathcal{M}} P_{jt}, \forall t \in \mathcal{S} \setminus \mathcal{M}$.
Append node $a^*$ to the route: 
$\mathcal{M} \leftarrow \mathcal{M} \cup \{ a^* \}$. 
\item Repeat steps \ref{item:mspa1}--\ref{item:mspa2} until the destination is added to the route.
\end{enumerate}
\end{algorithm}

\begin{remark}
Assuming that the value of the previous sum-of-received-power computations are cached, the complexity of step~\ref{item:mspa1} in MSPA is $O(D)$ because there are at most $(D-1)$ nodes not in the route.  The complexity of the comparisons in step \ref{item:mspa2} is $O(D)$.  Steps \ref{item:mspa1}--\ref{item:mspa2} are repeated at most $(D-1)$ times, giving a worst case complexity of the MSPA of $O(D^2)$. Recall that $D = |\mathcal{S}|$.
\end{remark}

We can show that the MSPA is optimal for DF on the phase fading Gaussian MRC.
\begin{thm}\label{thm:mspa}
In a Gaussian MRC in which the nodes send independent codewords, the MSPA route is optimal for DF.
\end{thm}

\begin{proof}[Proof for Theorem~\ref{thm:mspa}]
See \cite{ongmotani07secon} for proof.
\end{proof}

We now investigate how well the MSPA route $\mathcal{M}_{\text{MSPA}}$ performs compared to the optimal route in the Gaussian MRC. Due to the complexity involved in optimizing the power splits, we only simulate MRC up to 6 nodes, i.e., $D \leq 6$. For each $D$, we randomly place $D$ nodes in a $(D-1)\text{m}\times (D-1)\text{m}$ square area. The source and the destination are randomly chosen. We run the NNSA to find the optimal rate $R_{\text{DF}}^{\text{max}}$, and run the MSPA to find $R_{\text{DF}}(\mathcal{M}_{\text{MSPA}})$. The results are shown in Table~\ref{tab:mspa}. With high probability, the MSPA is able to find an optimal route. Also, $R_{\text{DF}}(\mathcal{M}_{\text{MSPA}})$ is a good indicator of $R_{\text{DF}}^{\text{max}}$.

\begin{table}[t]
\caption{Performance of the MSPA.}
\label{tab:mspa}
\centering
\begin{tabular}{| c  || c | c |}
\hline
$D$ & Average $R_{\text{DF}}(\mathcal{M}_{\text{MSPA}})/R_{\text{DF}}^{\text{max}}$ & Fraction of optimal routes using MSPA \\
\hline
3 & 1 & 1 \\
4 & 0.9999950 & 0.99882 \\
5 & 0.9999513 & 0.99522 \\
6 & 0.9999399 & 0.99194 \\
\hline
\end{tabular}
\end{table}

\begin{figure}[t]
\centering
\subfigure[A 4-node network]
{\includegraphics[width=0.25\linewidth]{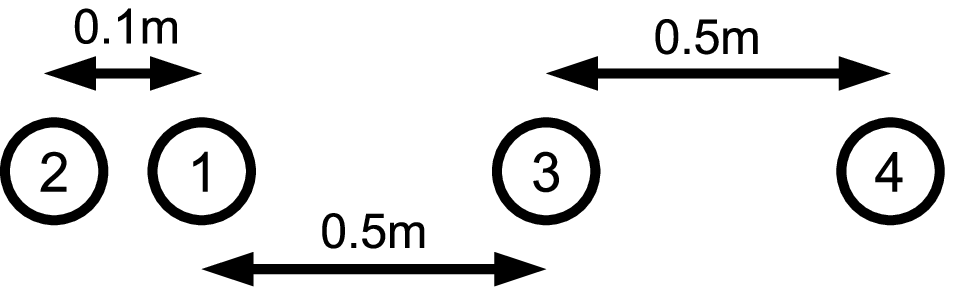}}
\subfigure[Information BER versus transmit SNR]
{\includegraphics[width=0.5\linewidth]{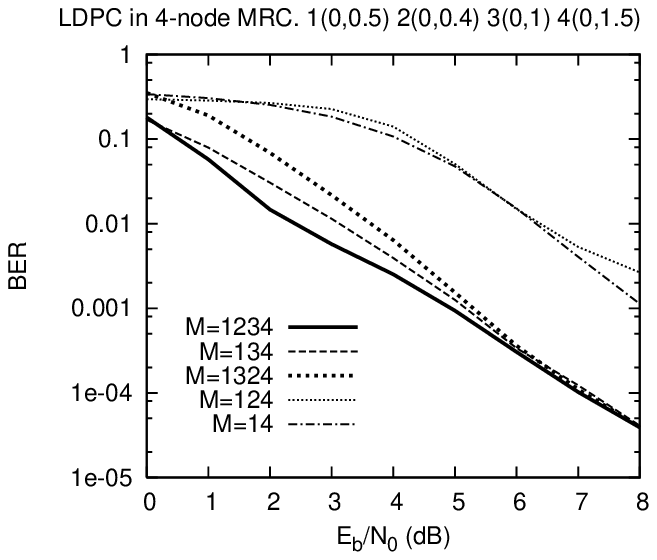}}
\caption{Performance of the LDPC coded DF strategy for different routes in the 4-node network.}
\label{fig:mh-4-nodes-ldpc}
\end{figure}

\section{DF with LDPC Codes} \label{sec:simulations}
In the previous section, we computed achievable rates of different routes using an information theoretic approach. In this section, we present the performance (via simulation) in a network using practical LDPC codes \cite{gallager62}\cite{mackay99} with incremental redundancy. In the literature, codes have been designed for the SRC~\cite{razaghiyu06}\cite{ezrigastpar06}\cite{khojastepouraazhang04}\cite{chakrabarti06}. Here, we implement a DF coding strategy on the MRC using rate-$\frac{1}{2}$ LDPC codes with puncturing and incremental redundancy.

Fig.~\ref{fig:mh-4-nodes-ldpc} gives the bit error rate (BER) versus transmit signal-to-noise ratio (SNR) for different routes. We find that different routes have significantly different performance (up to 4dB at BER=0.001 in this example). We also find that the NNSA route (also the MSPA route), i.e., $\mathcal{M}_{\text{MSPA}}= \mathcal{M}_{\text{DF}}^{\text{opt}} = \{1,2,3,4\}$, gives the best performance. Furthermore, in this setup, the NNSA/MSPA yields the optimal route directly, meaning that we do not need to evaluate all possible routes (5 routes in this example).

\begin{remark}
In practice, a relay in the route might decode a message wrongly and thus forward the wrong message. When this happens, the nodes behind, when trying to cancel the co-channel interference introduced by this relay, will introduce more noise at their decoders. While this scenario is not captured in \eqref{eq:df-rate}, we allow imperfect interference cancellation in all of our simulations here.
\end{remark}




\bibliography{bib}

\end{document}